\documentstyle[natbib,epsfig,longtable]{aipproc}

\newcommand{\mm}{ \mu^+\mu^-}

\begin{document}

\pagestyle{plain}
\footskip 1.5 cm

\title{Neutrino Radiation Challenges and Proposed Solutions
for Many-TeV Muon Colliders
\thanks{
To appear in Proc. HEMC'99 Workshop -- Studies on Colliders and
Collider Physics at the Highest Energies: Muon Colliders at 10 TeV to
100 TeV; Montauk, NY, September 27-October 1, 1999,
web page http://pubweb.bnl.gov/people/bking/heshop.
This work was performed under the auspices of
the U.S. Department of Energy under contract no. DE-AC02-98CH10886.}
}

\author{Bruce J. King}
\address{Brookhaven National Laboratory\\
email: bking@bnl.gov\\
web page: http://pubweb.bnl.gov/people/bking}
\maketitle

\begin{abstract}
 Neutrino radiation is expected to impose major design and
siting constraints on many-TeV muon colliders. Previous
predictions for radiation doses at TeV energy scales are
briefly reviewed and then modified for extension to the
many-TeV energy regime. The energy-cubed dependence of lower
energy colliders is found to soften to an increase of slightly
less than quadratic when averaged over the plane of the collider
ring and slightly less than
linear for the radiation hot spots downstream
from straight sections in the collider ring. Despite this,
the numerical values are judged to be sufficiently high that
any many-TeV muon colliders will likely be constructed
on large isolated sites specifically chosen to minimize or eliminate
human exposure to the neutrino radiation. It is pointed out
that such sites would
be of an appropriate size scale to also house future
proton-proton and electron-positron colliders at the
high energy frontier, which naturally leads to conjecture
on the possibilities for a new world laboratory for high
energy physics. Radiation dose predictions are also presented
for the speculative possibility of linear muon
colliders. These have greatly reduced
radiation constraints relative to circular muon colliders
because radiation is only emitted in two pencil
beams directed along the axes of the opposing linacs.
\end{abstract}

\section{Introduction}

\begin{figure}[t!] %
\centering
\includegraphics[height=2.3in,width=5.5in]{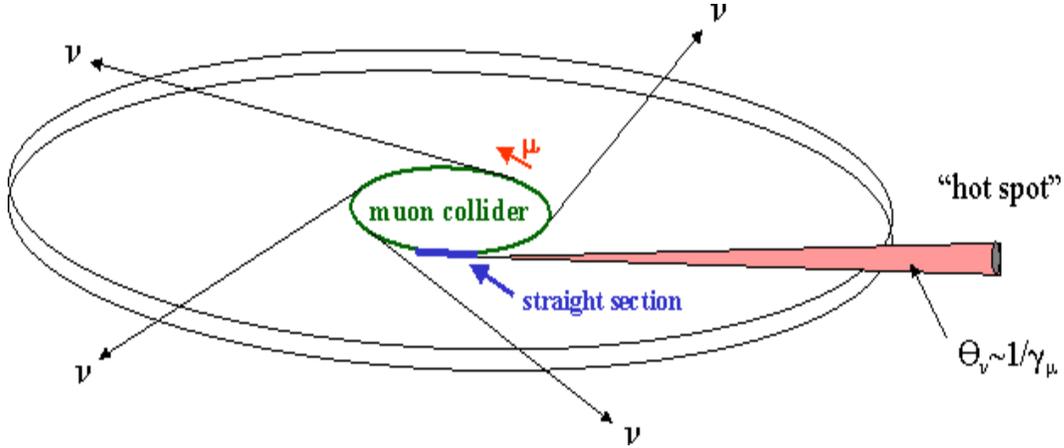}
\caption{
The decays of muons in a muon collider will produce a
neutrino radiation disk emanating out tangentially from the
collider ring. Radiation hot spots in the disk will occur
directly downstream from straight sections in the collider
ring.
}
\label{nurad_disk}
\end{figure}

 Neutrinos interact so rarely that, only
50 years ago, even their detection was not considered to
be feasible. It is therefore quite surprising that the
design of future muon colliders will usually be constrained by the need to
limit hazards from neutrino radiation~\cite{bjkthesis,pac99nurad}.
Neutrinos are produced
copiously at muon colliders from the decays of the
large currents of muons circulating in the collider ring:
\begin{eqnarray}
\mu^- & \rightarrow & \nu_\mu + \overline{\nu_{\rm e}} + {\rm e}^-,
                                             \nonumber \\
\mu^+ & \rightarrow & \overline{\nu_\mu} + \nu_{\rm e} + {\rm e}^+.
                                                 \label{nuprod}
\end{eqnarray}
The neutrino direction is tightly collimated to
within a characteristic angle, $\theta_\nu$, of the decaying
muon's direction, where:
\begin{equation}
\theta_\nu = 1/\gamma_\mu =
\frac{m_\mu c^2}{E_\mu} \simeq \frac{10^{-4}}{E_\mu [{\rm TeV}]},
                                                   \label{thetanu}
\end{equation}
for $\gamma_\mu$ the relativistic boost factor of the muon,
$m_\mu$ the muon rest mass, $c$ the speed of light and
$E_\mu$ the muon energy.
(Units are given in square
brackets in the equations throughout this paper.)
The combined effect of all the muon decays will be a disk of
neutrinos emanating out in the plane of the collider
ring~\cite{pac99nurad}, as
shown in figure~\ref{nurad_disk}. Straight
sections in the ring will cause radiation hot spots in the
disk~\cite{bjkthesis} where all of the decays in the straight
section line
up into a pencil beam that is superimposed on the disk, again
with a characteristic opening half-angle for the cone of
$1/\gamma_\mu$.
As a notable contrast to all other radiation hazards, the
neutrino attenuation length is too long for the beam to be
appreciably attenuated by any practical amount of shielding
material, including even
the expanse of ground between the collider ring and where
the radiation disk breaks ground.

  Example parameter sets for muon colliders that illustrate
the radiation hazard are given in table~\ref{tab:radvsE}.
The entries in the table will be referred to and explained
throughout this paper. For now, we note that the radiation
doses may be compared with the U.S. Federal off-site limit of
1 mSv/year or, in alternative units, 100 mrem/year.
(The limit is comparable to typical
background radiation levels of 0.4 to 4 mSv/year~\cite{PDG}.)
The radiation hazard is seen to rise sharply with
muon collider energy, increasing from a tiny fraction of the
legal limit for the lower energy colliders to well
above the limit for the collider scenarios at 10 TeV
and 100 TeV. This behavior will be quantified
in the two sections that follow; the next section will characterize
the radiation dose for colliders up to the TeV energy scale
and the section after that will discuss some mitigating factors
that come into play for the many-TeV muon colliders that are
the subject of
this workshop. Possible solutions for the hazard at many-TeV
energies will be discussed in the next-to-last section of
the paper before rounding out with a summary section.

\begin{table}[ht!]
\caption{Straw-man muon collider specifications and the
corresponding neutrino radiation parameters for muon colliders
at 0.1, 4, 100 and 100 TeV. The first two
muon collider scenarios were taken from references~\cite{status}
and~\cite{epac98}, respectively, while the final two scenarios
were straw-man parameter sets for this workshop~\cite{hemc99specs}.
The calculation of the neutrino radiation parameters is
discussed in the text.
}
\begin{tabular}{|r|cccc|}
\hline
\multicolumn{1}{|c|}{ {\bf center of mass energy, ${\rm E_{CoM}}$} }
                            & 0.1 & 4 TeV  & 10 TeV  &  100 TeV \\
additional description
                            & ${\rm H^0}$ factory & ``lite''
               & ${\rm 2^{nd}}$ gen. & ${\rm 3^{rd}}$ gen.  \\
collider luminosity, ${\cal L}$ [${\rm cm^{-2}.s^{-1}}$]
                                        & $1 \times 10^{31}$
                                        & $6 \times 10^{33}$
                                        & $1 \times 10^{36}$
                                        & $1 \times 10^{36}$ \\
collider int. lum., ${\int \cal L}$ [${\rm fb^{-1}/yr}$]
                                        & 0.1
                                        & 60
                                        & 10 000
                                        & 10 000 \\
\hline
muon beam energy, $E_\mu$ [TeV]         & 0.05 & 2 & 5 & 50 \\
muon decays/yr, $N^+_\mu\,[10^{20}]$   & 6 & 0.08 & 8 & 0.4 \\
collider reference depth, $d$ [m]      & 10 & 300 & 100 & 100 \\
$\nu$ beam distance to surface, $L$ [km]  & 11 & 62 & 36 & 36 \\
$\nu$ beam radius at surface [m]       & 24 & 3.3 & 0.8 & 0.08 \\
ave. rad. dose in plane [mSv/yr]       & $2 \times 10^{-5}$
                                       & $5 \times 10^{-4}$
                                       & 2.3 & 10  \\
str. sec. len. for 10x ave. rad. [m]
                                       & 1.9 & 1.1 & 1.0 & 4.2 \\
\hline
\end{tabular}
\label{tab:radvsE}
\end{table}

\section{The Neutrino Radiation Hazard for Muon Colliders
up to TeV Energy Scales}

 This section reviews reference~\cite{pac99nurad} in
characterizing the potential neutrino radiation hazard and giving numerical
estimates for the radiation dose in the so-called equilibrium
situation that pertains to the beams from muon colliders at
the TeV energy scale and below.

\begin{figure}[t!] %
\centering
\includegraphics[height=1.5in,width=6.0in]{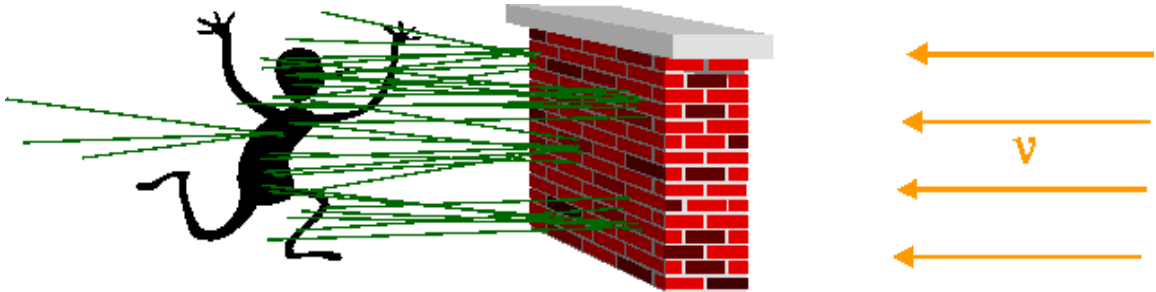}
\caption{
The radiation hazard arises primarily from particle showers
initiated by neutrino interactions in material near the
person.
}
\label{nurad_person}
\end{figure}

 Neutrinos at energies beyond a few GeV interact predominantly
through deep inelastic scattering off nucleons.
The radiation hazard arises from the showers of penetrating
charged particles produced through neutrino interactions with
any material bathed by the neutrino radiation disk, as is indicated
in figure~\ref{nurad_person}.
Starting from the initial interaction products,
an avalanche effect of secondary, tertiary, etc. interactions,
produces the vast majority of the charged particles. For TeV-scale
neutrinos, neutrino interactions in people themselves may therefore
only account for as little as 0.1\% of their radiation
dose~\cite{Mokhov_direct} because the primary hadrons
from the interaction will typically exit the person before interacting
to commence the shower of
charged particles.

 The development of particle showers makes the dose very
dependent on the local surroundings.
Radiation hazard calculations must conservatively consider
the worst case configuration where a person is (i) completely
bathed in the radiation disk or the pencil beam downstream
from a straight section
and (ii) is surrounded by material that will initiate
showers from neutrino interactions. A further requirement
for the validity of these calculations is that the characteristic
density of the surrounding material must be sufficient to
contain the showers within the pencil beam or radiation disk.
Such geometries can be contrasted with
the more common situation where the showers will
spread out transversely beyond the beam and, hence, dilute the
dose received by someone within the disk. Instead,
this containment criterion will be satisfied only by materials
where the nuclear interaction length that
characterizes shower development is shorter
than the beam radius, as will generally be the case
for solids or liquids but not for air or other
gaseous media. As examples~\cite{PDG}, water and quartz
have interaction lengths of 85 cm and 43 cm, respectively,
while the interaction length of air, 700 meters, is much
larger than the typical few-meter beam radii expected
at TeV-scale muon colliders (see table~\ref{tab:radvsE}).

\begin{figure}[t!] %
\centering
\includegraphics[height=2.3in,width=5.5in]{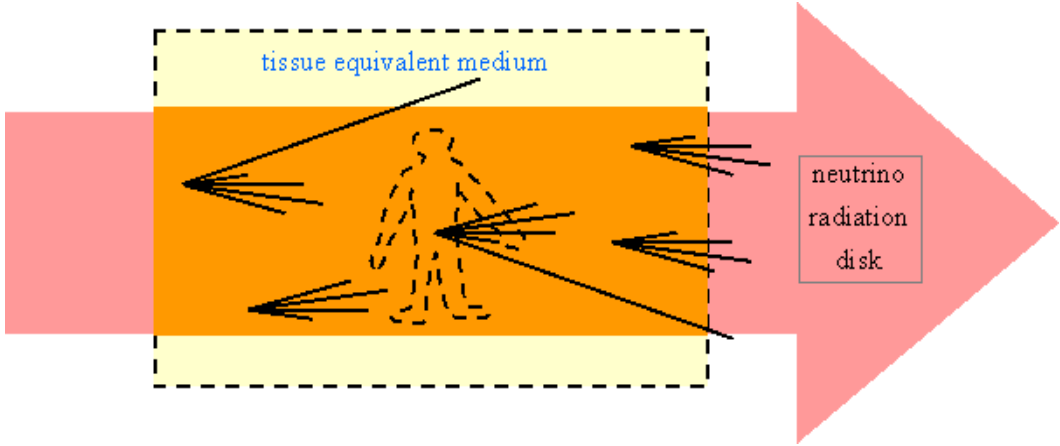}
\caption{
Conceptual illustration of the ``equilibrium approximation''
geometry used for quantitative worst-case radiation calculations.
The entire person is bathed by the neutrino radiation disk in
the plane of the collider ring and
perhaps additionally by the radiation cone downstream from a straight
section in the collider ring. The person is also enclosed in
a ``tissue equivalent medium'' that has sufficient density
to localize the charged particle showers from neutrino interactions
so that little of the showers' energy spills out beyond the
radiation disk.
}
\label{nurad_equil}
\end{figure}

 The worst-case situation can be conveniently modeled
for numerical calculations~\cite{pac99nurad} by
considering a person completely enclosed within a ``tissue
equivalent medium'' i.e. a medium with the approximate density
of water. (The ``scuba diver'' configuration.) This geometry
is illustrated in figure~\ref{nurad_equil}.
For this simplified model and rather artificial geometry, it is
clear that the energy--per--unit--mass
absorbed by a person is constrained simply by conservation of
energy to be approximately equal to the summed energy of
the neutrino interactions in the person. This applies even though most
of the deposited energy is from shower products of interactions
upstream from the person. This approximate equality is referred
to as the equilibrium approximation.
For more realistic and general geometries with inhomogeneous
distributions of mass, it can be argued~\cite{pac99nurad}
that the
equilibrium approximation is either valid or, alternatively,
conservatively overestimates the radiation dose.

 The calculation of radiation doses is straightforward~\cite{pac99nurad}
within this equilibrium assumption because 
neutrino interaction
cross sections are well known and the approximate neutrino flux
within the pencil beams can be simply predicted from the known
decay kinematics and relativistic kinematics. Here we merely reproduce, in a slightly reorganized form, the formula of
reference~\cite{pac99nurad} for the average whole-body
radiation dose, $D^{ave}$, in the plane of the collider ring:
\begin{equation}
D^{ave} [mSv] \simeq 3.7 \times
N^+_\mu[10^{20}] \times \frac{(E_\mu[TeV])^3}{(L[km])^2}.
                           \label{dosevalue}
\end{equation}
$N^+_\mu$ is the number of muons decaying in the collider
ring, per year and per charge sign and given
in appropriate units of $10^{20}$ decays. $L$ is the tangential
distance from the ring to where the dose is measured --
typically where the radiation disk exits the ground.

  The additional radiation hot-spot from a straight section of
length $l^{ss}$ is given by~\cite{pac99nurad}:
\begin{equation}
D^{ss} [mSv] = 1.1 \times 10^5 \times N_\mu[10^{20}] \times f^{ss} \times 
 \frac{(E_\mu[TeV])^4}{(L[km])^2},
                                \label{fssdose}
\end{equation}
where the fraction, $f^{ss}$, of the
ring circumference, $C$, corresponds to $l^{ss}$ through:
\begin{equation}
f^{ss} = \frac{l^{ss}}{C}.
  \label{fss}
\end{equation}
Equation~\ref{fssdose} can be rewritten in terms of $l^{ss}$ and
the average bending magnetic field in the collider ring, $B^{ave}$, as
\begin{equation}
D^{ss} [mSv] \simeq 5.3 \times
N^+_\mu[10^{20}] \times l^{ss}[m] \times
B^{ave}[T] \times
 \frac{(E_\mu[TeV])^3}{(L[km])^2},
                           \label{ssdose}
\end{equation}
by making use of the relation
\begin{equation}
C[km] = \frac{2 \pi \cdot E_\mu[TeV]}{0.3 \cdot B^{ave}[T]}.
   \label{Bave}
\end{equation}

 Equations~\ref{dosevalue},~\ref{fssdose} and~\ref{ssdose} are not claimed
to be accurate at much better than order-of-magnitude
level and detailed follow-up Monte Carlo simulations have confirmed
their predictions to this level of accuracy~\cite{Mokhov_MC}.

 The ratio of equations~\ref{dosevalue} and~\ref{ssdose}
immediately gives the length of straight section,
$l^{equiv}$, that approximately doubles the in-plane average
radiation dose:
\begin{equation}
l^{equiv} [{\rm meters}] \simeq \frac{0.7}{B_{ave}[T]}.
                                    \label{lequiv}
\end{equation}
This is only of order 10 cm for the typical average bending fields
expected in muon collider storage rings.

 The energy-cubed dependences of equations~\ref{dosevalue}
and~\ref{ssdose} deserve further comment. The dependence
in equation~\ref{dosevalue} is the product of three linear
factors: 1) the inverse width of the disk, which goes as
$1 / \theta_\nu$, 2) the neutrino cross-section, $\sigma_{\nu N}$,
and 3) the average neutrino energy, $<E_\nu >$, that is deposited
per interaction:
\begin{equation}
{\rm disk\;average\;dose},\;D^{ave} \sim \frac{1}{\theta_\nu} \cdot
             \sigma_{\nu N} \cdot <E_\nu>
            \;\propto\; E_\mu \cdot E_\mu
            \cdot E_\mu = E_\mu ^3.
  \label{avescaling}
\end{equation}
For the straight sections, the inverse disk width is replaced
by the inverse cross-sectional area of the pencil beam, which goes
as $1 / \theta_\nu^2$. Also, the proportionality of
equation~\ref{fssdose} on $f^{ss}$ brings in a factor of
$1/E_\mu$ for a given value of $l^{ss}$ in equation~\ref{ssdose},
using equations~\ref{fss} and~\ref{Bave}. The energy scaling
of equation~\ref{ssdose} can therefore be broken down into:
\begin{equation}
{\rm str.\;section\;dose},\;D^{ss} \sim (\frac{1}{\theta_\nu})^2 \cdot
             \sigma_{\nu N} \cdot <E_\nu> \cdot f^{ss}
            \;\propto\; E_\mu^2 \cdot E_\mu
            \cdot E_\mu / E_\mu = E_\mu ^3.
  \label{ssscaling}
\end{equation}

  Equations~\ref{dosevalue} and~\ref{ssdose} predict the
numerical radiation doses in table~\ref{tab:radvsE} for the 0.1 TeV and
4 TeV collider examples.
 The values used for $L$ assume a collider
located at a specified depth, $d$, under a site with a smooth surface
having the average curvature of the Earth, so that
\begin{equation}
L_{exit} = \left( 2 \times d \times R_E \right) ^{1/2},
    \label{Lfromd}
\end{equation}
where the Earth's radius has the value
$R_E = 6.4 \times 10^6$ m and the equation very reasonably assumes
that $d \ll R_E$.

 The preceding discussion in this section has provided the background
information for the numerical estimates of table~\ref{tab:radvsE}
for the 0.1 TeV and 4 TeV parameter sets.
The tabulated radiation predictions for the 10 TeV and 100 TeV
collider parameters anticipate some mitigating factors at
multi-TeV energies that will now be discussed.

\section{Mitigating Factors at Many-TeV Energies}

\begin{figure}[t!] %
\centering
\includegraphics[height=2.0in,width=6.0in]{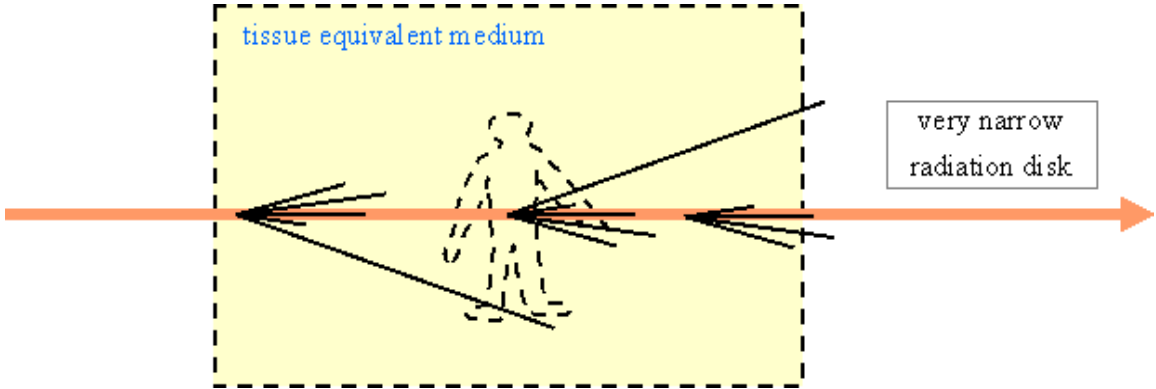}
\caption{
The worst-case geometry for radiation exposure at many-TeV muon
colliders, to be contrasted with the ``equilibrium approximation''
geometry of the preceding figure.
}
\label{nurad_nonequil}
\end{figure}

\begin{table}[ht!]
\caption{
Parameterization of the cross section factor $X(E_\mu)$ that
describes the fall-off from a linear cross section rise with
energy. By definition, $\alpha \equiv \log_{10}(E_\mu[TeV])$. The
parameterizations are simple logarithmic energy interpolations
between the cross sections given in reference~\cite{quigg}.
The numerical factors in the expressions, 1.453, 1.323, 1.029,
0.512 and 0.175 are the total summed neutrino-nucleon and
antineutrino-nucleon cross-sections-divided-by-energy~\cite{quigg}
at the neutrino energies of 100 GeV, 1 TeV, 10 TeV, 100 TeV and 1 PeV,
respectively, given in units of $10^{-38}\;{\rm cm^2/GeV}$.
As an adequate approximation to avoid convolutions with neutrino
energy spectra, the muon energies in the table have been set
equal to the corresponding neutrino energies.
By definition, $X(E_\mu) \equiv 1$ for $E_\mu = 100$ GeV,
which is the reference energy used for the radiation dose
parameterizations of equations~\ref{dosevalue} and~\ref{ssdose}.
}
\begin{tabular}{|c|c|}
\hline
muon energy range  &   expression for $X(E_\mu)$  \\
\hline
$E_\mu<1$ TeV             &
                 $(-1.453\times\alpha+1.323\times(\alpha+1))/1.453$ \\
1 TeV$<E_\mu<$ 10 TeV     &
                 $(1.323\times(1-\alpha)+1.029\times\alpha)/1.453$  \\
10 TeV$<E_\mu<$ 100 TeV   &
                 $(1.029\times(2-\alpha)+0.512\times(\alpha-1))/1.453$ \\
100 TeV$<E_\mu<$ 1000 TeV &
                 $(0.512\times(3-\alpha)+0.175\times(\alpha-2))/1.453$ \\
1000 TeV$<E_\mu$          &
                 $(0.175/1.453) \times 3^{3-\alpha}$ \\
\hline
\end{tabular}
\label{tab:Xvalues}
\end{table}

 Equations~\ref{dosevalue} and~\ref{ssdose} are too pessimistic
at the many-TeV energies
addressed at this workshop, for two reasons. The smaller of the
two effects is a partial leveling off
in the neutrino cross section. Rather than
a continuation of the linear rise up to TeV energy scales, it is
predicted that~\cite{quigg}, for example, the neutrino cross
section at 100 TeV
is only 33 times that at 1 TeV, instead of a 100-fold increase.

  More significantly, the beam radius, $\theta_\nu L$, ceases
to be large compared to the size of a person or to the width of the
shower it produces.
This prevents the possibility of a person ever experiencing the
geometry of figure~\ref{nurad_equil} that is needed for the
equilibrium approximation to apply. Instead, a modified worst-case
geometry that often applies for many-TeV muon colliders is
illustrated in figure~\ref{nurad_nonequil}.

The formulae~\ref{dosevalue} and~\ref{ssdose}
can be modified to apply roughly to the case of a narrow, many-TeV 
neutrino radiation disk or pencil beam
by considering how the equilibrium approximation breaks down
for a beam radius that, due to either increasing $E_\mu$
or decreasing $L$, becomes comparable to or smaller than a hadron
shower radius. Instead of depositing the
dose over a progressively smaller vertical band (in the case of
a radiation disk), or a spot (in the case of the pencil beam),
the shower radius imposes a limiting transverse size scale.
For definiteness, the
calculations for table~\ref{tab:radvsE} spread the disk-average
dose out evenly over a half-height of 0.5 meters -- which is
comparable to the 0.43 m
interaction length of granite that was mentioned previously --
and spread the hot-spot dose over a circle of the same radius.

  The explicit many-TeV equations corresponding to the lower
energy equations~\ref{dosevalue} and~\ref{ssdose} become,
respectively:
\begin{equation}
D^{ave}_{many-TeV} [mSv] \simeq 3.7 \times
N^+_\mu[10^{20}] \times  \frac{(E_\mu[TeV])^3}{(L[km])^2}
                 \times  X(E_\mu) \times F(E_\mu, L)
                           \label{highEdosevalue}
\end{equation}
and
\begin{eqnarray}
D^{ss}_{many-TeV} [mSv] \simeq 5.3 \times
N^+_\mu[10^{20}] \times l^{ss}[m] \times
B^{ave}[T] \times
 \frac{(E_\mu[TeV])^3}{(L[km])^2} \nonumber \\
                  \times  X(E_\mu) \times (F(E_\mu, L))^2,
                           \label{highEssdose}
\end{eqnarray}
where  $X(E_\mu)$ and $F(E_\mu, L)$ are the high energy
suppression factors for the cross section leveling and
small spot size, respectively.

  The cross-section suppression factor, $X(E_\mu)$, has, by definition,
the value unity for $E_\mu=100$ GeV, which was the energy chosen to
calculate the numerical coefficients for equations~\ref{dosevalue}
and~\ref{ssdose}. Its slow decrease with increasing energy has
been approximated by simple logarithmic energy interpolations
between the cross sections given in reference~\cite{quigg} and
the form and coefficients for the interpolations are given in
table~\ref{tab:Xvalues}.

  The appropriate form of $F(E_\mu, L)$, incorporating the assumed
minimum transverse shower dimension of 0.5 m that was discussed
above, is:
\begin{equation}
F(E_\mu, L) =  {\rm min \left(1,
         \frac{\theta_\nu[rad]L[km]}{5 \times 10^{-4}}\right)}
 =  {\rm min \left(1,
         \frac{L[km]}{5 \times E_\mu [TeV]}\right)},
  \label{Fspot}
\end{equation}
where equation~\ref{thetanu} has also been used to obtain the
second form of the expression.

  Substituting in the explicit form of equation~\ref{Fspot}
for the case where $L[km] < 5 \times E_\mu [TeV]$ returns
the narrow-beam form of
equations~\ref{highEdosevalue} and~\ref{highEssdose}:
\begin{equation}
D^{ave}_{many-TeV} [mSv] \rightarrow 0.74 \times
N^+_\mu[10^{20}] \times  \frac{(E_\mu[TeV])^2}{L[km]}
                 \times  X(E_\mu)
                           \label{limhighEdosevalue}
\end{equation}
and
\begin{equation}
D^{ss}_{many-TeV} [mSv] \rightarrow 0.21 \times
N^+_\mu[10^{20}] \times l^{ss}[m] \times
B^{ave}[T] \times E_\mu[TeV] \times  X(E_\mu).
                           \label{limhighEssdose}
\end{equation}
The second of these equations is independent of the distance $L$
in this limit, as it intuitively must be: the beam is now narrow
enough that a person will intercept essentially the entire beam, so
the dose should become independent of distance in this limit.

 We can now review the energy scaling of the radiation dose at
many-TeV energies, to compare the power law dependences with
those for the lower energy colliders that are given in
equations~\ref{avescaling} and~\ref{ssscaling}.
The 0.5 m fixed half-height removes the
dependence of equation~\ref{avescaling} on the disk height,
$\theta_\nu L$. Combined with the partial
leveling of the cross-section, this softens the radiation
rise with energy relative to slightly less than quadratic
in energy:
\begin{equation}
{\rm many\!-\!TeV\;disk\;average\;dose},\;D^{ave}_{many-TeV} \sim 
             \sigma_{\nu N} \cdot <E_\nu>
            \;\propto\;  E_\mu^{<1}
            \cdot E_\mu = E_\mu^{<2}.
  \label{highEavescaling}
\end{equation}
Similarly, the many-TeV dose from straight sections loses both powers
of $E_\mu$ that came from the $1/(\gamma_\mu L)^2$ factor in
equation~\ref{ssscaling}, giving:
\begin{equation}
{\rm many\!-\!TeV\, str.\, sec.\,dose},\;D^{ss}_{many-TeV} \sim
             \sigma_{\nu N} \cdot <E_\nu> \cdot f^{ss}
            \;\propto\; E_\mu^{<1}
            \cdot E_\mu / E_\mu = E_\mu^{<1},
  \label{highEssscaling}
\end{equation}
i.e. a less-than-linear rise with energy for radiation
hot spots from a straight section of fixed length $l^{ss}$.

  The assumed shower radius of 0.5 meters is clearly a somewhat
arbitrary choice, so it should be borne in mind that
equations~\ref{highEdosevalue} and~\ref{highEssdose}
will be even more approximate than the lower energy predictions
of equations~\ref{dosevalue} and~\ref{ssdose}. These predictions
at many-TeV energies should also be interpreted even more as
worst case scenarios than those at lower energies, since the
material surrounding the person is now required to have a
density comparable to granite in order to confine the hadron showers 
to the assumed transverse dimensions of 0.5 meters.

 No detailed 
follow-up Monte Carlo simulations for various material geometries
have yet been performed for the Many-TeV scenarios discussed in
this section. Such simulations would be very valuable in confirming and
refining the rough numerical estimates obtained from applying the
above high-energy modifications to
equations~\ref{dosevalue} and~\ref{ssdose}.

\section{Proposed Solutions}

\begin{figure}[t!] %
\centering
\includegraphics[height=4.5in,width=4.5in]{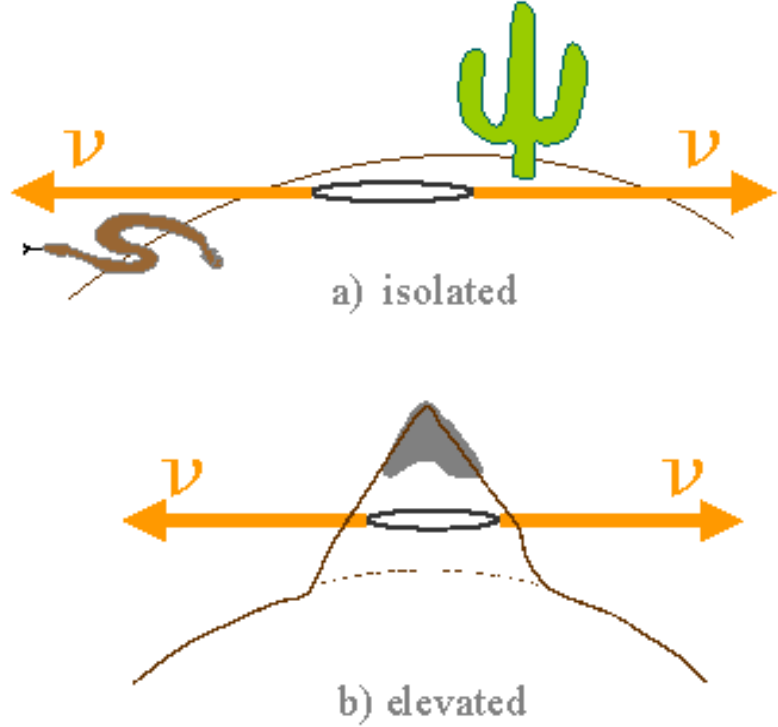}
\caption{
Two site options that avoid exposing people to the neutrino radiation
disk in the plane of the muon collider ring.
}
\label{nurad_site_options}
\end{figure}

\begin{figure}[t!] %
\centering
\includegraphics[height=3.0in,width=3.37in]{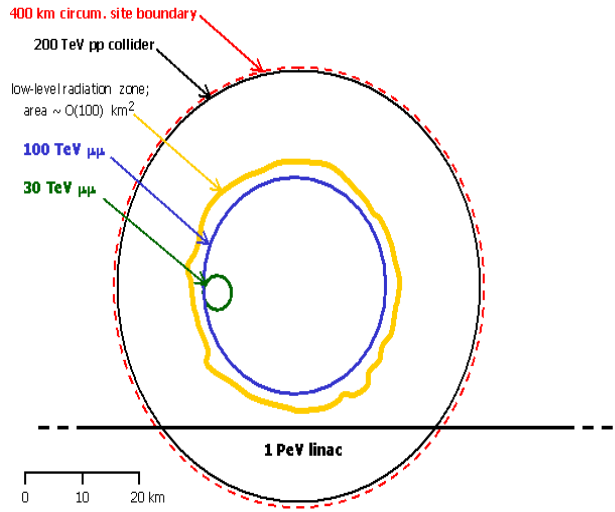}
\caption{
The ultimate high energy physics laboratory ? See further
discussion in reference~\cite{hemc99intro}.
Figure reproduced from reference~\cite{hemc99intro}.
}
\label{worldlab}
\end{figure}

\begin{figure}[t!] %
\centering
\includegraphics[height=3in,width=3in]{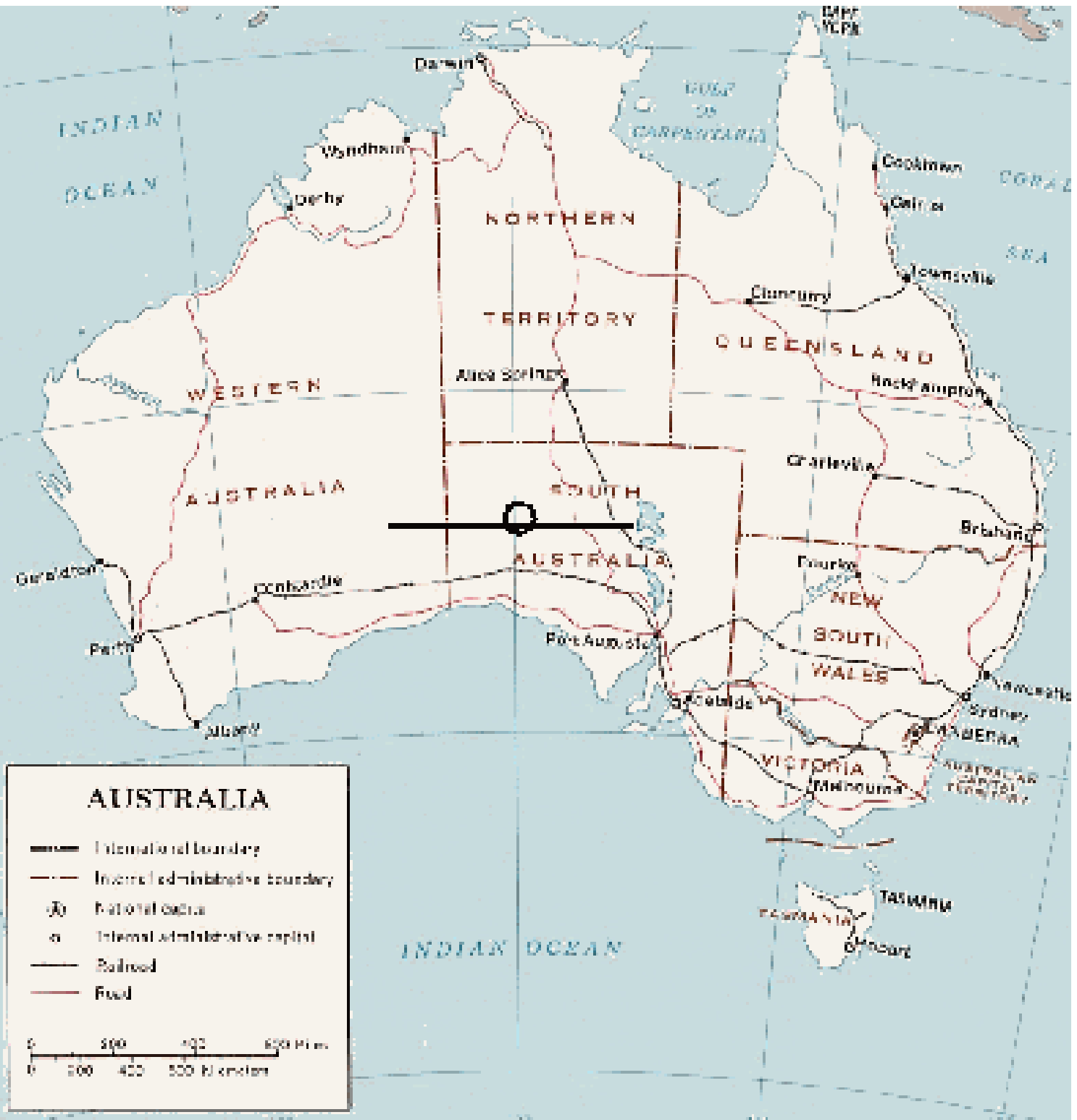}
\caption{
  An example to illustrate the size of a HEP laboratory
with a 400 km site boundary circumference. A circle of this
diameter has been drawn in the Great Victoria Desert
(just above the ``A'' in the label ``South Australia''),
showing that the outline of a laboratory of this size
would even be visible on a map of Australia.
A 1000 km long strip of land for the 1 PeV linear collider
would perhaps not be included inside the original laboratory
boundary. The choice of country and positioning of the site
are for illustration only.
Figure reproduced from reference~\cite{hemc99intro}.
}
\label{labAustralia}
\end{figure}

  Several means to reduce the radiation hazard have been
proposed previously in reference~\cite{pac99nurad}:
\begin{enumerate}
  \item  minimize straight sections in the collider ring, e.g. by
superimposing some bending field on all focusing magnets
  \item  improve the luminosity per unit current, from better
beam cooling etc.
  \item  use fenced-off radiation enclosures downstream from
the largest straight sections
  \item  bury the muon collider ring deeper underground to increase
the distance before the neutrino disk exits the ground. Optionally,
the ring can also be tilted and oriented to take advantage of
natural geological features
  \item  choose to build the muon collider on a site where human
exposure to the radiation disk will be minimized or, ideally,
nobody at all will be exposed to the neutrino radiation.
\end{enumerate}

  The orders-of-magnitude reductions desired for many-TeV
colliders appear difficult to achieve through any combination
of items 1 through 4 alone so
the final option -- a specially chosen site -- may well
be unavoidable for muon colliders at the highest energies.

 Two classes of siting options are shown in figure~\ref{nurad_site_options}.
The first is an isolated site, where nobody is exposed to the
radiation disk before it exits into the atmosphere due to the
local curvature of the Earth. The height above 
ground, $h$, at distance $L$ from a
collider ring close to the surface of a spherical Earth
is given by rearranging equation~\ref{Lfromd} to:
\begin{equation}
h = \frac{L^2}{2 R_E}.
   \label{hfromL}
\end{equation}
As an example that is discussed further below,
a distance to the site boundary of $L=64$ km
corresponds to $h=$ 320 m, which might be considered a very comfortable
clearance height at the site boundary for an isolated region.
Some zoning restrictions could also
be placed on tall structures near the laboratory site
if this was additionally required.

 The second option in figure~\ref{nurad_site_options}, siting
the laboratory on elevated land, would take advantage of
the local topology to provide either a smaller site or higher
clearances for the radiation disk at the site boundary. In
practice, both siting options would likely be combined by
choosing the most elevated site in an isolated region.

 A third siting option, placing the collider in a valley or other
depressed region to extend the distance before the disk exits the
ground, is discussed elsewhere in these proceedings~\cite{Colin}.
This can be considered to be a variation on item 4 of the
above list. As speculation, this siting option might well
be practical up to perhaps the 10 TeV energy scale but would
likely give an inadequate dose reduction for muon colliders
at the highest energies and luminosities.

  The transient radiation doses to people in planes
and birds flying through the disk would clearly be negligible
for either of the siting options of
figure~\ref{nurad_site_options}
since the doses in table~\ref{tab:radvsE} must be
accumulated over a full accelerator year of $10^7$ seconds
and also, as explained previously, the conservatively calculated
doses are anyway about three orders
of magnitude above the full-year doses for a person or bird
when they are not surrounded by dense material.

  Because of the dilute beam halo from showers induced in the atmosphere,
the radiation dose at the site boundary would not be strictly
zero even if the neutrino radiation disk was well above ground
level. Speculatively, the halo might
extend down below the neutrino disk by perhaps of order the
interaction length of air, $\lambda_{\rm air} = 700$ m.
On naively comparing with equations~\ref{Fspot}
and~\ref{highEdosevalue},
the disk average dose might
be expected to be down by of order (0.5 m)$/\lambda_{\rm air}$,
which is about three orders of magnitude, from the in-disk
prediction of equation~\ref{highEdosevalue}. Similarly, comparison
with equation~\ref{highEssdose} suggests that the additional
doses from straight sections might speculatively be predicted
to be diluted by the square of this ratio, i.e. by about
six orders of magnitude.
(These predictions are essentially just repeating the
arguments leading up to equations~\ref{limhighEdosevalue}
and~\ref{limhighEssdose}, but now with a limiting radial extent
of 700 meters rather than 0.5 meters.)
If these tentative predictions are valid then both the average
dose and the straight section dose would be safely below
radiation limits. However, the predictions are very
speculative and detailed Monte Carlo showering calculations
are called for to predict the true level of the beam halo for
this geometry.

 On the positive side, a site diameter of order 100 km
is also an appropriate size scale for jointly housing the largest
potentially plausible proton or electron colliders along with the
many-TeV muon colliders. This is the case in the
``proof-of-plausibility'' scenario for future energy frontier
colliders that is presented in reference~\cite{hemc99intro}
and that might lead to a site layout that is illustrated
conceptually in figure~\ref{worldlab}. As a detail on
the site layout, the various muon acceleration and collider
rings could plausibly be placed in the same plane to minimize
the radiation zone to perhaps of order 100 square kilometers.

  Just to illustrate the size scale, figure~\ref{labAustralia} shows
a circle with a 400 km circumference (i.e. a 64 km radius) drawn in
an unpopulated desert
region of Australia. The content of both
figures~\ref{worldlab} and~\ref{labAustralia} is further
discussed in reference~\cite{hemc99intro}. Note that the figures
also include a linear collider, whose radiation hazards we now discuss.

\section{Neutrino Radiation at Many-TeV Linear Muon Colliders}

\begin{figure}[t!] %
\centering
\includegraphics[height=3.0in,width=4.5in]{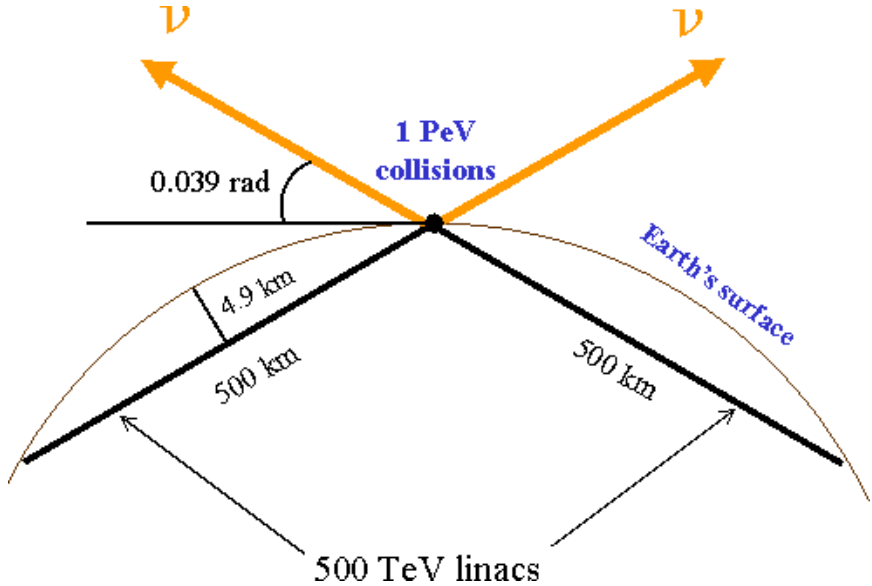}
\caption{
 A PeV-scale linear muon collider~\cite{Zimmermann} would shoot two
neutrino beams upwards and towards the sky at angles to the horizontal
of perhaps a few tens of milliradians, depending on the length
of the linacs.
}
\label{nurad_1PeV}
\end{figure}

  It was speculated during this workshop that the 
ultimate potential energy reach for muon colliders
could be extended by using linear $\mm$ colliders,
and straw-man parameter sets were presented for such
single pass colliders~\cite{Zimmermann}.

  From the standpoint of neutrino radiation, a linear muon
collider has two advantages over circular muon colliders.
Firstly, the radiation is confined to two pencil beams which
would naturally be oriented at an upward tilt of perhaps
tens of milliradians, as illustrated in figure~\ref{nurad_1PeV}.
Secondly, the spent muons after each single-pass collision
can be immediately ranged out in a beam dump rather than
surviving until they decay into high energy neutrinos. This gives
a large reduction in the radiation dose within the pencil
beams.

  We now derive the dose for the unpleasant and very artificial
situation of a person living full-time in dense material immediately
downstream from a linac, in the center of one of the neutrino beams.
This will then be used to assess doses for more realistic situations.

  The calculation proceeds by considering the total dose as an
integral over the dose contributions, $\delta D(E)$, received in
each energy interval, $[E,E+dE]$, as the muon bunch accelerates
to the beam energy, $E_\mu$:
\begin{equation}
D^{linac} = \int_{0}^{E_\mu} dE\, \delta D(E).
   \label{doseint}
\end{equation}
We already have the means to estimate $\delta D(E)$, since
it can be compared to the dose from a straight section in a
circular collider. After scaling by the relative fraction
of muons decaying in the two cases, one obtains:
\begin{equation}
\delta D(E) =  D^{ss}_{many-TeV}(E)
                 \times \frac{1}{f^{ss}}
                 \times \frac{df}{dE}(E)\, dE,
\end{equation}
where $\frac{df}{dE}(E)\, dE$ is the fraction
of muons that decay in the energy interval
$[E,E+dE]$. This is given by:
\begin{equation}
\frac{df}{dE}(E) = \frac{1}{ \gamma_\mu \beta c \tau \times g},
  \label{df}
\end{equation}
where
the product $\gamma_\mu \beta c \tau = 660$ is the characteristic
decay length of ultra-relativistic muons,
with $\beta c \tau = 660$ meters, and
$g = dE/dz$ is the acceleration gradient.


Substituting in the expressions and constants from
equations~\ref{limhighEssdose}, ~\ref{thetanu}, ~\ref{fss} and~\ref{Bave}
gives, after some algebra and on taking care with units:
\begin{equation}
D^{linac}[mSv] = \frac{0.67 \times N_\mu[10^{20}]} {g[GeV/m]}
    \times \int_{0}^{E_\mu[TeV]} dE\, E\, X(E).
   \label{Dlinac}
\end{equation}
To solve this equation analytically, it is an adequate approximation to
replace the energy-weighted integral of $X(E)$
by its value at, say, $E=E_\mu/2$, giving finally:
\begin{equation}
D^{linac}[mSv] \sim \frac{0.33\, N_\mu[10^{20}]} {g[GeV/m]}
    \times  X(E_\mu/2)  \times (E_\mu[TeV])^2.
   \label{Dvaluelinac}
\end{equation}

  As a numerical example, the straw-man parameters for the 1 PeV muon
linear collider of reference~\cite{Zimmermann} assume
$E_\mu = 500$ TeV and $N_\mu = 0.064 \times 10^{20}$ in a
$10^7$ second year. Substituting in these values, with
the additional assumption that $g=1$ GV/m, gives:
\begin{equation}
D^{linac}{\rm \:[2 \times 500\:TeV\:example]} = 1400\;{\rm mSv/year}.
\end{equation}
To put this figure for a whole-year dose in perspective, it is
approximately 30 times the recommended maximum dose for a U.S.
radiation worker,
i.e. 50 mSv/year~\cite{PDG}, so such a radiation worker would be able to
work directly in the beam for of order 100 hours of accelerator
running time. This shows that, while the dose would be well above
legal limits for long-term occupancy, any doses from transient
exposure would still be relatively small. In particular, the calculation
gives reassurance on the negligible dose that would be received
by a bird or plane flying through the beam.

%





\section{Summary}

  Neutrino radiation is a very serious problem and design
constraint for muon colliders, particularly at very high
center-of-mass energies. A characterization of the neutrino
hazard has been presented that quotes the numerical formulae from
reference~\cite{pac99nurad} applying for TeV energy scales and
below, and that extends the numerical predictions up to many-TeV
energies.

 It has been shown that the radiation dose in the plane of the
muon collider ring rises quickly as the cube
of the beam energy up to around the TeV collider energy scale
before leveling off for many-TeV colliders to a slightly less
than quadratic rise with energy on average and slightly less
than linear for the radiation hot spots downstream from a
fixed length of straight section.

  To solve the neutrino radiation problem, many-TeV muon colliders
and their associated neutrino radiation disks may be located
within a new world laboratory
with site diameter of order 100 km and located at a carefully
chosen site in either the U.S., Canada, Australia, Northern
Europe or elsewhere where large isolated tracts of land can
be found with resources for a large-scale high technology
laboratory.

\end{document}